\documentclass{egpubl}
\usepackage{eg2025}
 
\ShortPresentation      
\usepackage[T1]{fontenc}
\usepackage{dfadobe}  

\usepackage{cite}  
\BibtexOrBiblatex
\electronicVersion
\PrintedOrElectronic
\ifpdf \usepackage[pdftex]{graphicx} \pdfcompresslevel=9
\else \usepackage[dvips]{graphicx} \fi

\usepackage{egweblnk}

\usepackage{array}
\usepackage{amsmath}
\usepackage{amssymb}
\usepackage{bm}
\usepackage{xcolor}
\usepackage{booktabs}
\usepackage{subcaption}
\usepackage{caption}
\title{\scalebox{0.9}{Pixels2Points: Fusing 2D and 3D Features for Facial Skin Segmentation}}
\author[V.\,Y. Chen \textit{et al.}]
{\parbox{\textwidth}{\centering
        V.\,Y. Chen$^{2}$\orcid{0009-0000-3353-3768},
        D. Wang$^{1}$\orcid{0000-0002-2879-6114},
        S. Garbin$^{1}$\orcid{0009-0000-5005-8110},
        J. Bednarik$^{1}$\orcid{0000-0002-4018-4954},
        S. Winberg$^{1}$\orcid{0000-0003-1066-8955},
        T. Bolkart$^{1}$\orcid{0000-0002-3829-3924},
        T. Beeler$^{1}$\orcid{0000-0002-8077-1205}
        }
        \\
{\parbox{\textwidth}{\centering $^1$Google $^2$ETH Z\"urich}
}
}

\begin{document}

\teaser{
 \includegraphics[width=500px]{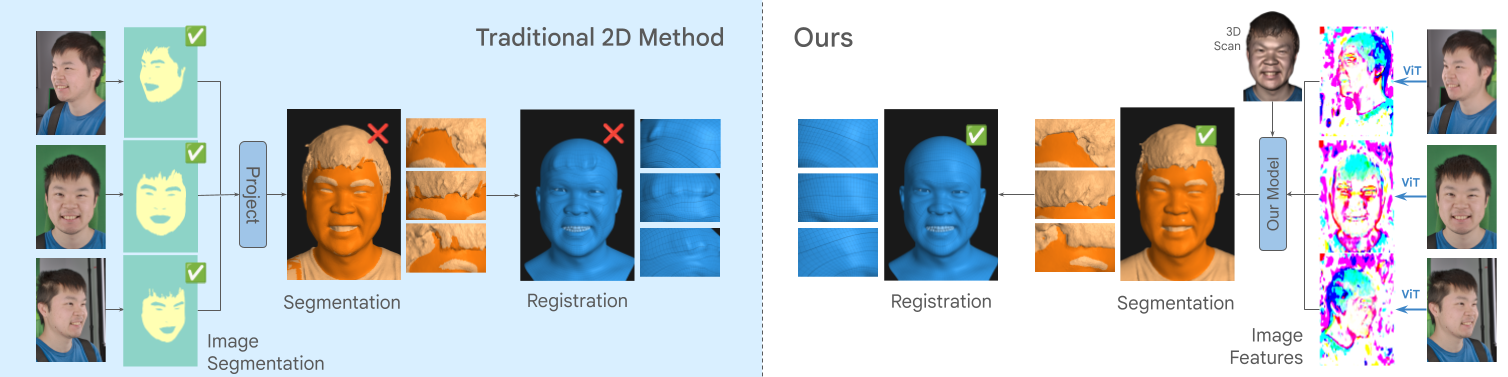}
 \centering
\caption{
 Human head registration aligns a template head mesh to the skin region of a 3D head scan. Previous methods first segment images and subsequently project 2D segmentation mask onto the 3D scan to differentiate between skin and non-skin areas. However, 3D scans are susceptible to reconstruction artifacts like interconnected hair, which are not present in the 2D image space. These artifacts persist in the 3D skin mask, leading to inaccurate registration results. Our model uses both the 3D scan and image ViT features to output per-vertex labels directly, resulting in a clean skin/non-skin separation and thus better registration.}
\label{fig:teaser}
}

\maketitle
\begin{abstract}

Face registration deforms a template mesh to closely fit a 3D face scan, the quality of which commonly degrades in non-skin regions (e.g., hair, beard, accessories), because the optimized template-to-scan distance pulls the template mesh towards the noisy scan surface.
Improving registration quality requires a clean separation of skin and non-skin regions on the scan mesh. 
Existing image-based (2D) or scan-based (3D) segmentation methods however perform poorly.
Image-based segmentation outputs multi-view inconsistent masks, and they cannot account for scan inaccuracies or scan-image misalignment, while scan-based methods suffer from lower spatial resolution compared to images.
In this work, we introduce a novel method that accurately separates skin from non-skin geometry on 3D human head scans.
For this, our method extracts features from multi-view images using a frozen image foundation model and aggregates these features in 3D.
These lifted 2D features are then fused with 3D geometric features extracted from the scan mesh, to then predict a segmentation mask directly on the scan mesh. 
We show that our segmentations improve the registration accuracy over pure 2D or 3D segmentation methods by 8.89\% and 14.3\%, respectively.
Although trained only on synthetic data, our model generalizes well to real data.


\end{abstract}  
\section{Introduction}

Face meshes in dense semantic correspondence play a crucial role in a wide spectrum of computer graphics applications, ranging from face animation~\cite{beeler2011high} to statistical model building~\cite{li2017learning}. 
The registration process to get 3D face meshes in correspondence commonly follows a two-step process: multi-view stereo (MVS) reconstruction and mesh registration~\cite{egger20203d}. 
MVS reconstructs an unstructured scan mesh from calibrated multi-view images, and the registration then deforms a template mesh to closely fit the skin region of this scan mesh. 
One key challenge lies in the presence of extraneous elements in the input scan such as hair, accessories, scanning artifacts, etc., which result in bumps and artifacts in the registration mesh surface (see Fig.~\ref{fig:teaser}).

To separate noise from signal, it is essential to segment the skin region from the rest of the 3D scan.
%
%
While image segmentation has achieved significant progress~\cite{kirillov2023segment,sarkar2023parameter}, projecting 2D labels onto a 3D surface presents several challenges. 
Multi-view inconsistency and the discrepancy between raw scan and images cause the projected segmentation masks to be suboptimal~(Fig.~\ref{fig:teaser}). 
%
Reconstruction artifacts are also inherently invisible to 2D segmenters as they only appear on the 3D surface.
Alternatively, pure scan-based 3D segmentation methods~\cite{sharp2022diffusionnet} face the challenges of low-resolution and irregular 3D geometry, compared to the dense regular data that image segmenters can leverage.

In this paper, we propose to use the best of both worlds by combining lifted 2D image features and 3D geometry features as input to a 3D model to segment the scan into skin and non-skin regions.
We extract image features from the multi-view input images with DinoV2~\cite{oquab2023dinov2}, project them into the 3D space, and fuse the features across views \cite{bolkart2023instant}.
The lifted 2D features are then combined with geometric features \cite{ovsjanikov2009hks,pauly2002efficient} extracted from the scan surface, and input to a DiffusionNet-based model \cite{sharp2022diffusionnet} to predict the segmentation masks.
%
The model is trained in a fully-supervised manner on synthetic data~\cite{wood2021fake} and generalizes well to real data.
%

Various techniques have been proposed for mapping 2D features onto 3D for segmentation, with most focusing on resolving multi-view inconsistencies in 2D image maps. 
Solutions range from attention-based mechanisms \cite{robert2022learning, jain2024odin} to depth-guided merging \cite{kundu2020virtual}. 
%
However, these methods overlook the inherent geometrical properties of the 3D surface, which provide crucial information for accurately segmenting unwanted regions.
Also, existing work focuses on scene segmentation, the task which is less sensitive to segmentation errors on a very fine level; on the other hand, a highly accurate vertex mask is essential in human head segmentation. To the best of our knowledge, our work is the first one to tackle facial segmentation by fusing 2D and 3D features.

\vspace{-0.2cm}

\section{Methodology}

\begin{figure}[t]
  \centering
  \includegraphics[width=.45\textwidth]{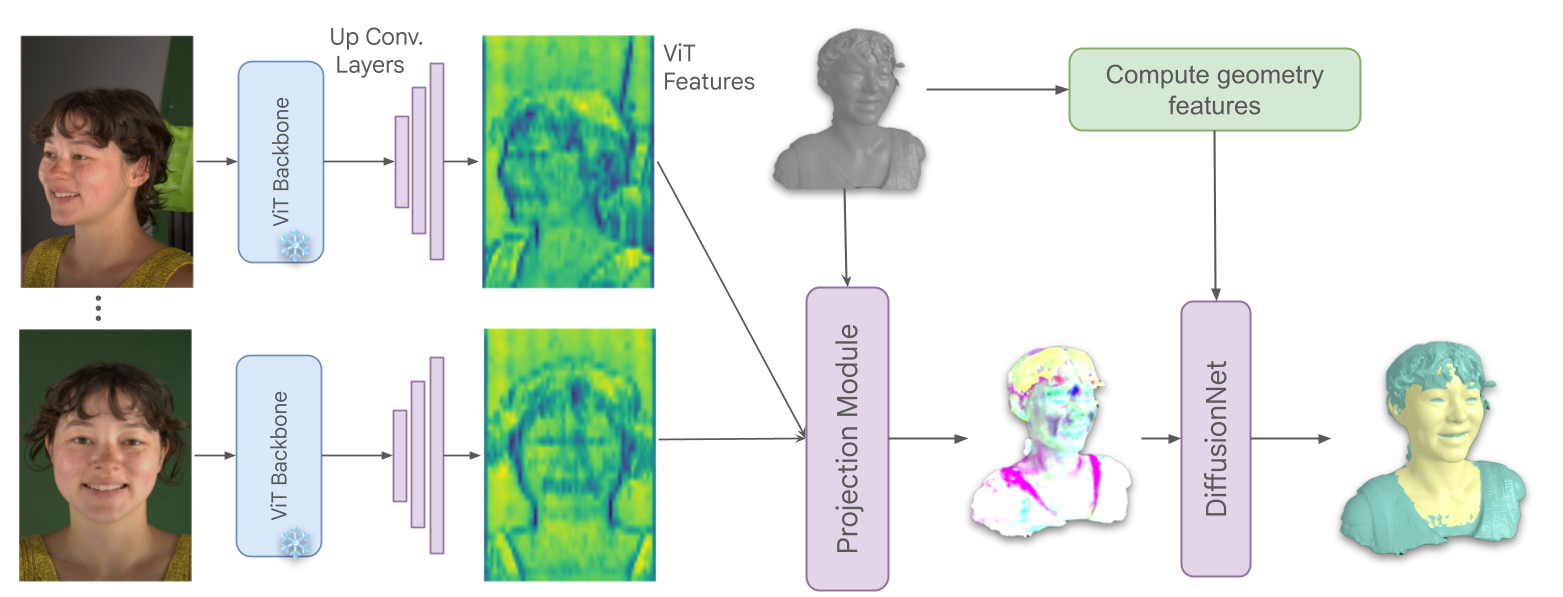}
  \caption{
  \label{fig:model-overview}
  \textbf{Architecture overview.} Given multi-view images of an expressive human face and a reconstructed 3D scan, we first extract features from images using a frozen ViT model and upscale them back to image resolution using up-convolution layers. Then, with precomputed camera parameters, we project the features onto the mesh vertices and fuse features across views weighted by vertex visibility. A combination of geometric and ViT features is fed into DiffusionNet, outputting the final segmentation.}
  \vspace{-0.5cm}
\end{figure}

Given multi-view images $I_{n}, n \in \{1, ..., N\}$ for $N$ views, each with camera extrinsic and intrinsic parameters, and a reconstructed scan mesh, our method outputs segmentation labels per scan vertex, separating skin from non-skin regions (Fig.~\ref{fig:model-overview}).

\vspace{-0.2cm}

\subsection{Feature Extraction}

\paragraph*{Image Features.}
For each $I_{n}$, we use  DinoV2~\cite{oquab2023dinov2} to extract image features $F_{n} = \text{head}(\text{ViT}(I_{n}))$.
Features from multiple internal ViT layers are concatenated and a ViT $\text{head}$~\cite{ranftl2021vision} is applied to reassemble ViT tokens and upsample the feature back to the image resolution. 

We project each scan vertex $\bm{v}\in \mathbb{R}^3$ into the image space of each view with the perspective projection determined by the camera parameters and bilinearly sample the corresponding feature map at the projected vertex location.
This gives us feature vectors $\mathbf{f}_{n}, n \in \{1, ..., N\}$ for each view.
Inspired by TEMPEH~\cite{bolkart2023instant}, we then fuse the features across all views using visibility-weighted mean $\bm{\bar{\mu}} = \sum_{n=1}^{N} w_n \mathbf{f}_n$ and variance $\bm{\bar{\sigma}}^{2} = \sum_{n=1}^{N} w_n (\mathbf{f}_n - \bm{\mu})^2$, where $w_n$ is the normalized weight based on the vertex visibility from the given camera. 
%
Features from vision foundation models are multi-view consistent~\cite{el2024probing}, so clean and smooth mesh regions will have similar features across different views. On the contrary, regions with reconstruction artifacts will therefore have high feature variance across different views (see Sup. Mat.). 

\paragraph*{Geometric Features.} Per-vertex geometric features are used to complement ViT features. We employ Heat Kernel Signature (HKS)~\cite{ovsjanikov2009hks}, and surface variation ($\sigma_{30}$)~\cite{pauly2002efficient}. The latter is less susceptible to noise and performs well in ML-based 3D surface analysis tasks compared to other geometric features like curvature~\cite{gundogdu22garnetplusplus} (see Sup. Mat.).

\vspace{-0.2cm}

\subsection{Scan Segmentation Model}\label{methodology:3d-segment}

%
We use DiffusionNet~\cite{sharp2022diffusionnet} as the 3D segmentation network, which leverages spatial diffusion along surfaces to enable communication between vertices. The diffusion process does not rely on a specific mesh representation, making DiffusionNet inherently agnostic to changes in mesh resolution or geometric representations.
DiffusionNet accounts for the underlying shape of the mesh and effectively propagate properties across vertices.
The network takes per-vertex feature as input and output per-vertex segmentation label. It is trained using cross-entropy loss. 

\vspace{-0.2cm}

\subsection{Synthetic Data Generation}
\label{methodology:training-data}

\begin{figure}[tb]
\begin{center}
\small
\setlength{\tabcolsep}{1pt}
\newcommand{\height}{1.8cm}
\begin{tabular}{ccc}
  \includegraphics[height=\height]{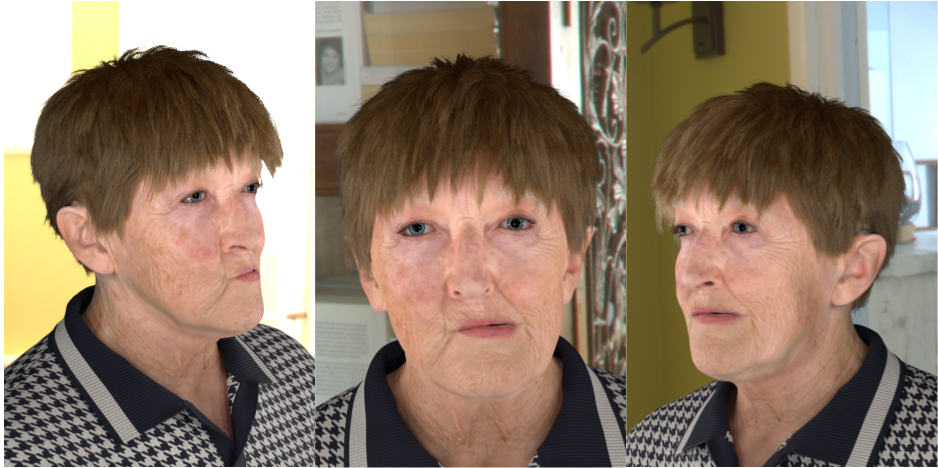}  
  & \includegraphics[height=\height]{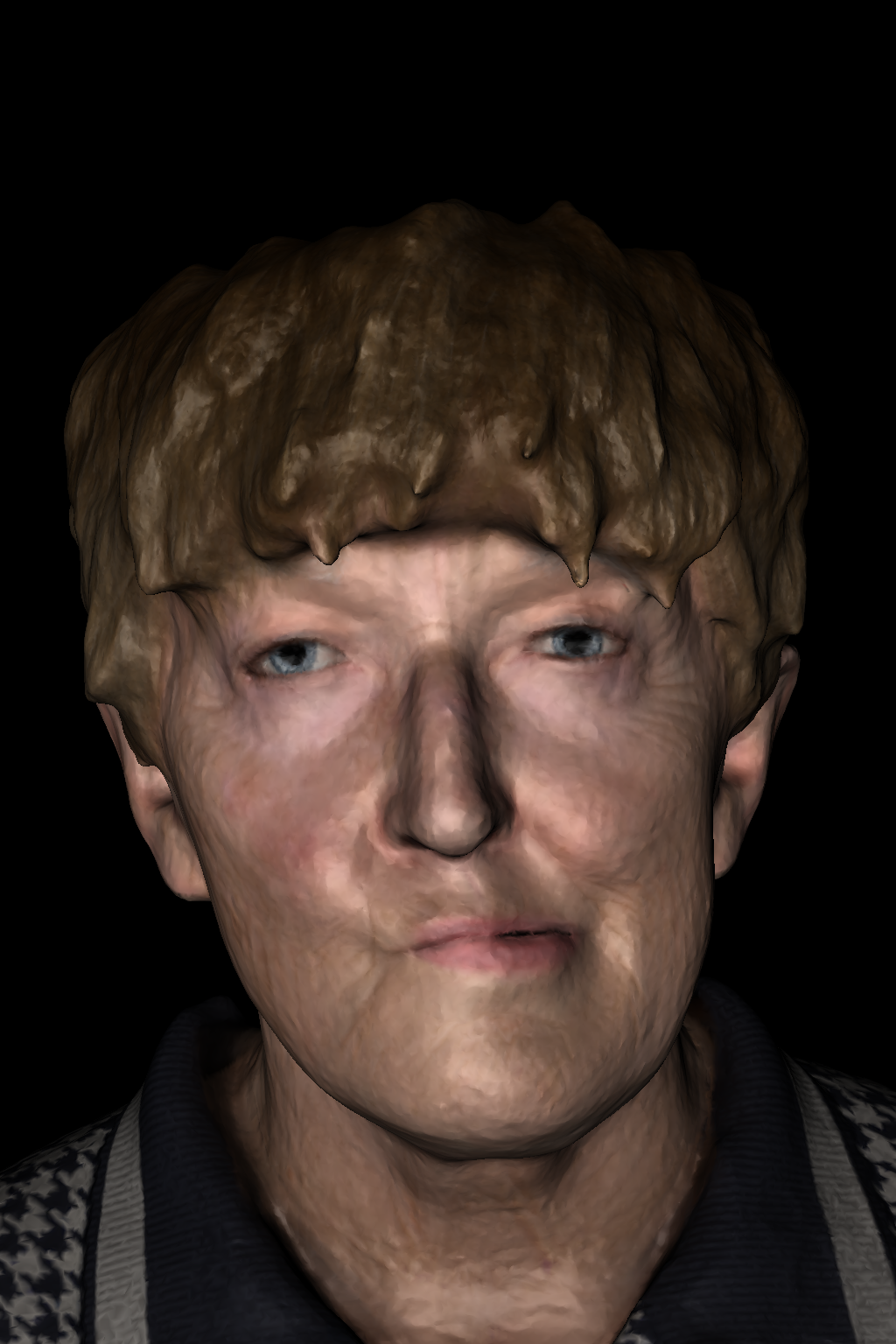}  
  & \includegraphics[height=\height]{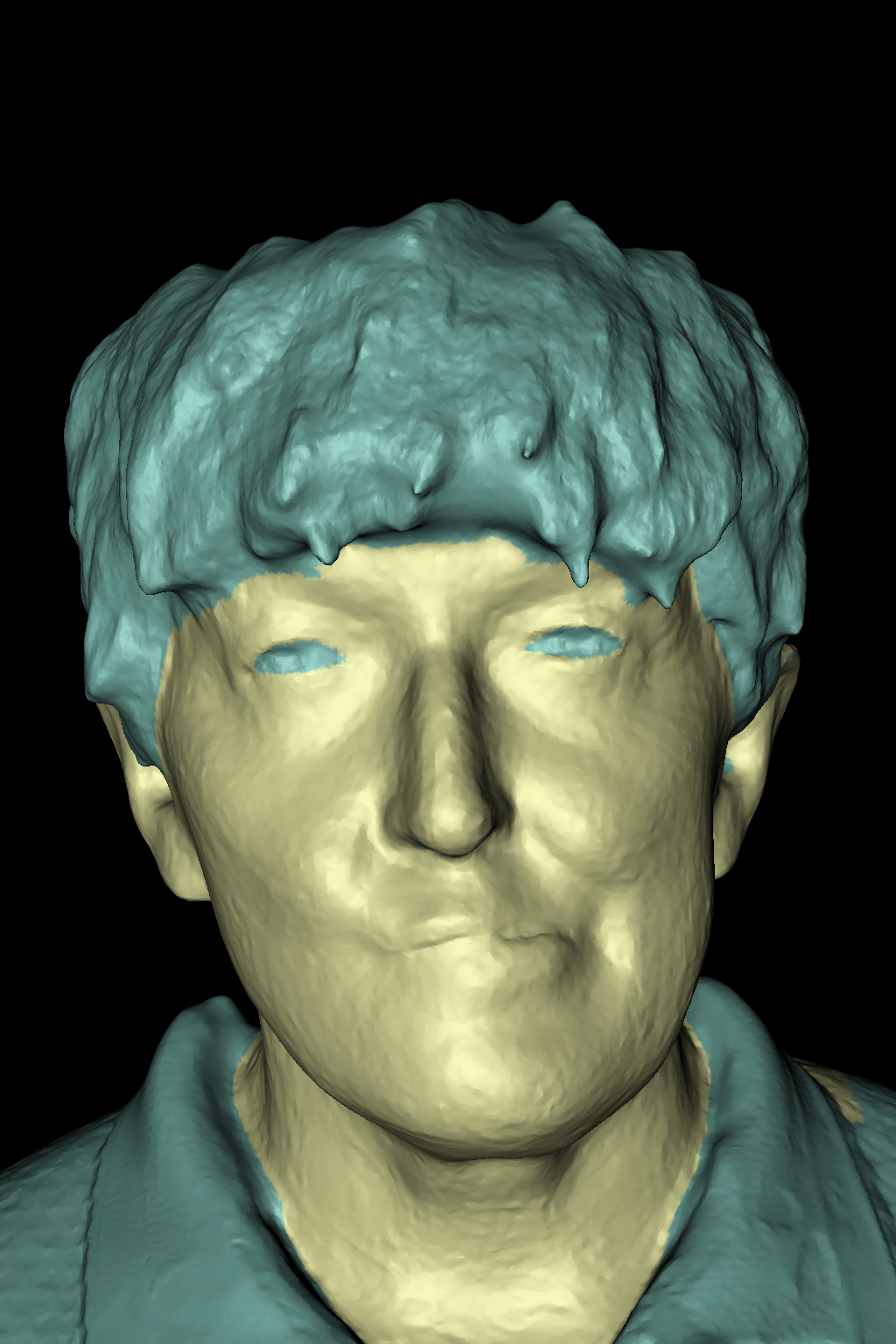} 
\\
  \includegraphics[height=\height]{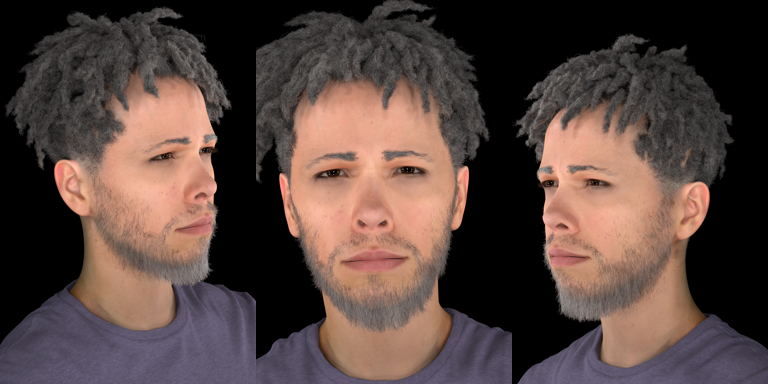}  
  & \includegraphics[height=\height]{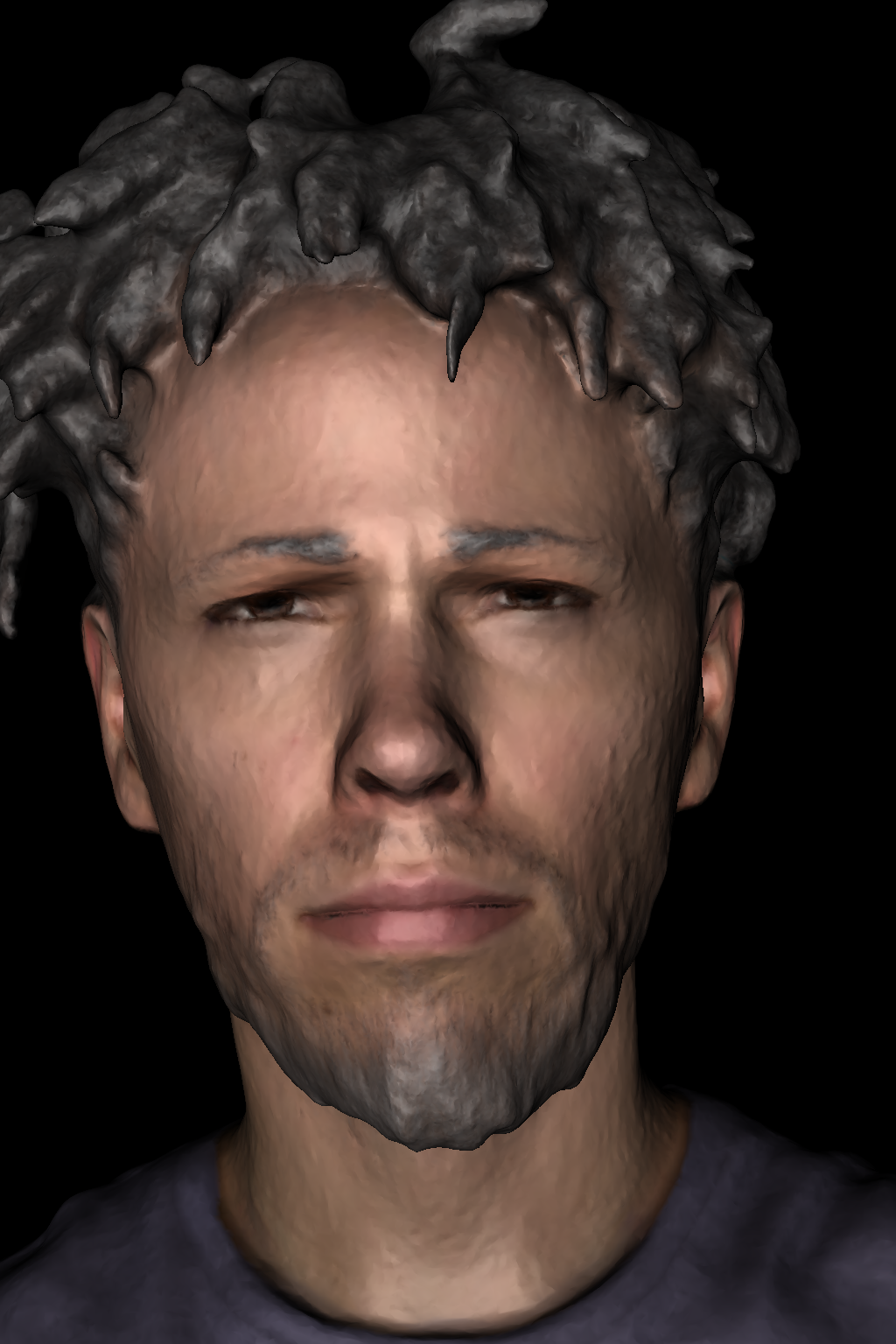}  
  & \includegraphics[height=\height]{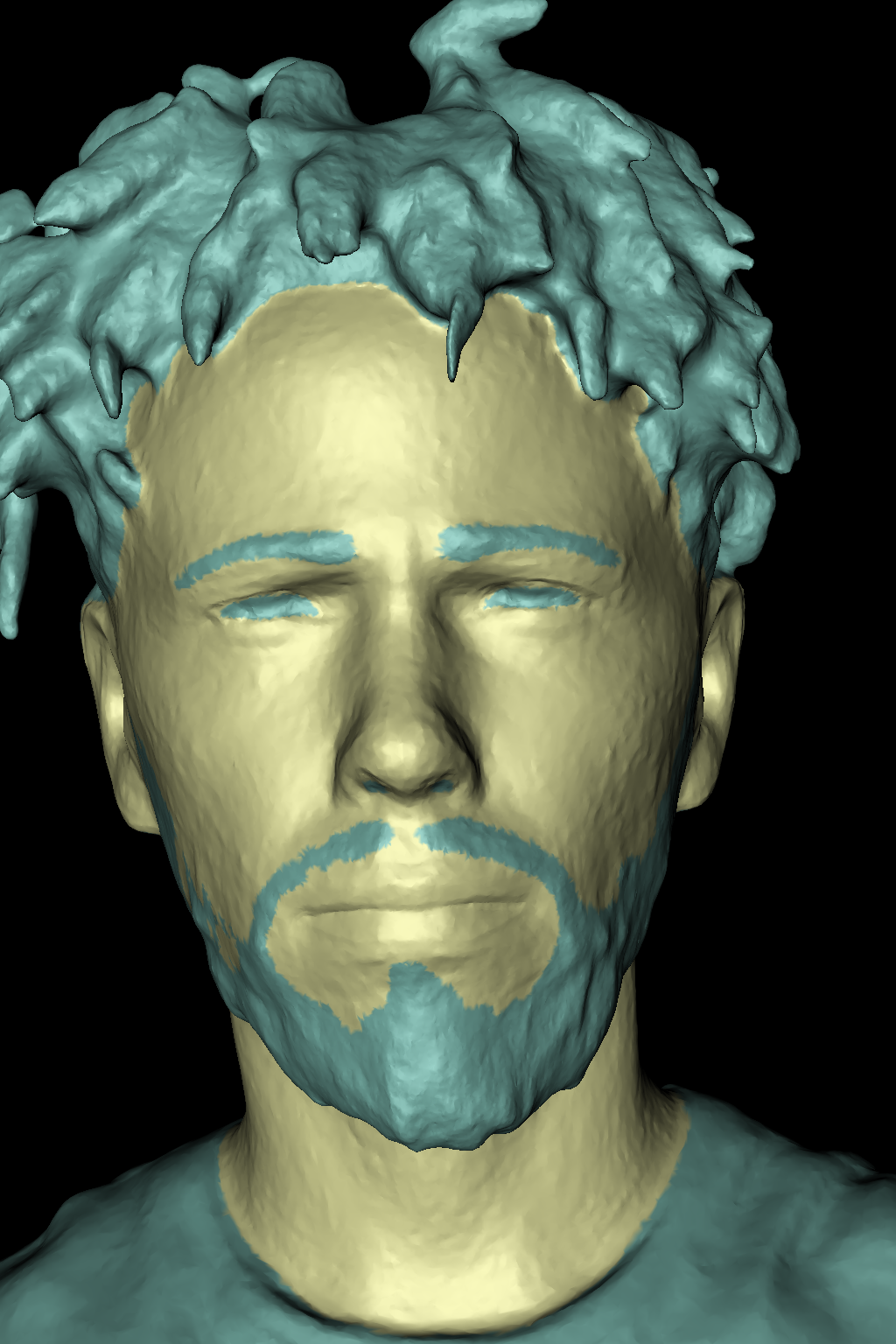} 
\\
Rendering & Reconstruction & Label
\end{tabular}
\end{center}
\caption{\label{fig:synthetic-data} \textbf{Synthetic data}. From left to right: images of synthetic human head rendered from fixed views (3 out of 13); reconstructed scan mesh using MVS and Poisson surface reconstruction; synthetic ground truth segmentation labels on the scan mesh.}
\vspace{-10pt}
\end{figure}

We train exclusively on synthetic 3D head scans (Fig.~\ref{fig:synthetic-data}, more in Sup. Mat.). Following the Procedural Human approach~\cite{wood2021fake}, we first generate the underlying 3D head geometry by sampling identity and expression from a custom 3DMM model (see Sup. Mat.). Then we attach skin textures, hairstyles, and clothing assets sampled from our digital asset library. Subsequently, the 3D head is placed in a synthetic scene with a randomly sampled background. We render the scene from 13 fixed views using Blender Cycles.
We then reconstruct 3D scan meshes from the rendered images using multi-view stereo \cite{qiu2024chosen}, where each scan consists of approximately 250K vertices.
To get ground truth segmentation labels, we compute the point-to-surface distance from each scan mesh vertex to the underlying 3DMM head mesh, and label vertices with a distance smaller than 1.5 mm as skin region, and the remaining vertices as non-skin region.
In total, we generate 3000 synthetic head scans and split these into 2500 training, and 500 test examples.

\vspace{-0.2cm}
\section{Results}\label{sec:results}
In this section, we compare our method to pure 2D and 3D methods, and provide ablation experiments investigating the impact of the ViT backbone, the projection method, and the input features. 

\paragraph*{Evaluation Metric.}
The segmentation quality is quantified with the Mean Intersection-over-Union (mIoU).
While the synthetic data has ground truth segmentation labels, for real capture data, we manually annotate the skin region of 20 scans (see Sup. Mat. for the labeling protocol).
The downstream registration quality is measured through the surface-to-point distance ($d_{\text{surface}}$) from registration results to the underlying ground truth facial skin surface (the "spa mask" region, see Sup. Mat.). 
The 3DMM template mesh in the procedurally-generated human head serves as the ground truth for synthetic data, while for real data, we use the manually annotated skin region of the scans as the pseudo ground truth.

\vspace{-0.2cm}

\subsection{Comparison with 2D and Pure 3D Methods}
\begin{figure*}[ht]
\begin{center}
\small
\setlength{\tabcolsep}{2pt}
\newcommand{\height}{2.3cm}
\begin{tabular}{ccccccc}
  \includegraphics[height=\height]{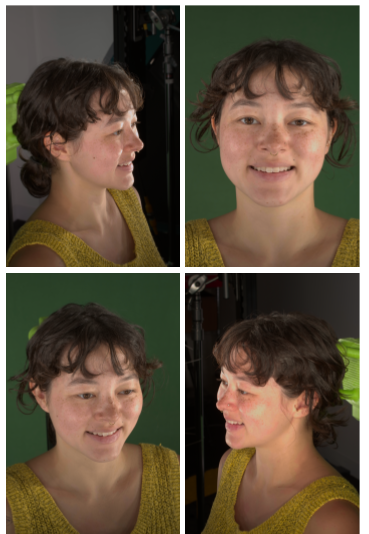}  
  & \includegraphics[height=\height]{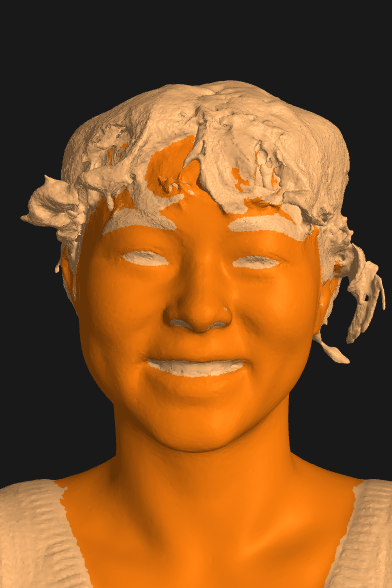}  
  & \includegraphics[height=\height]{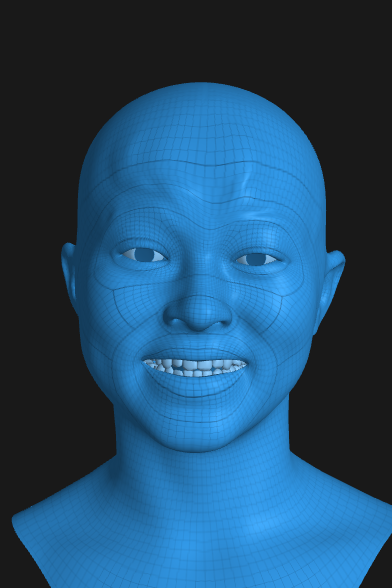} 
  & \includegraphics[height=\height]{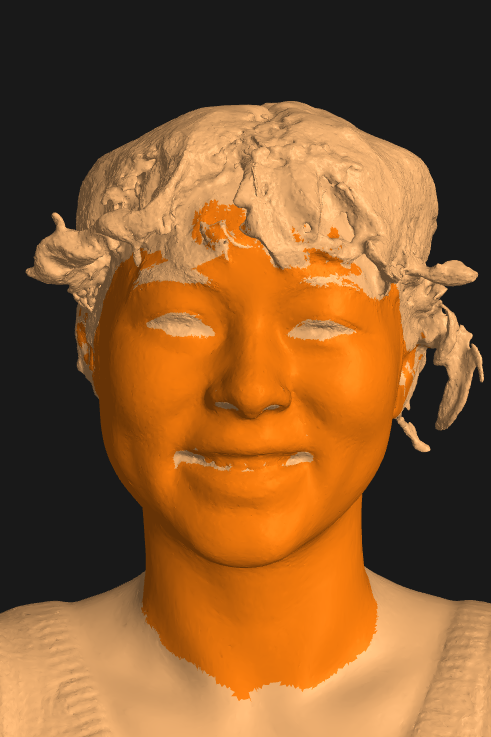} 
  & \includegraphics[height=\height]{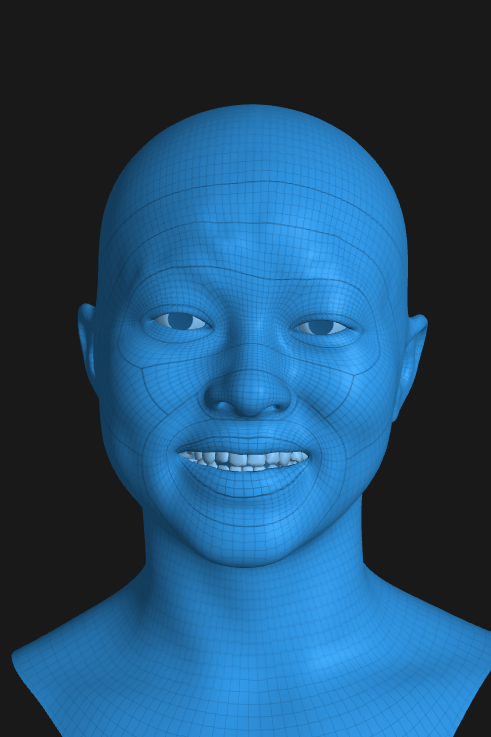} 
  & \includegraphics[height=\height]{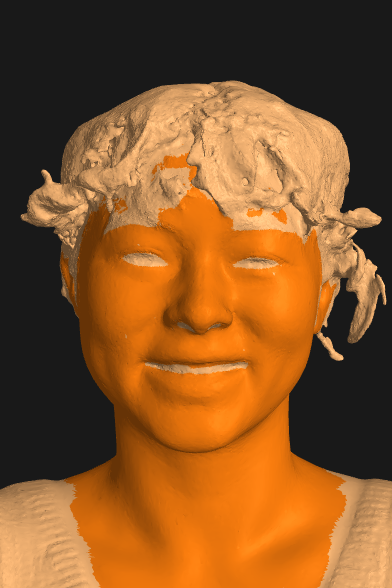}
 & \includegraphics[height=\height]{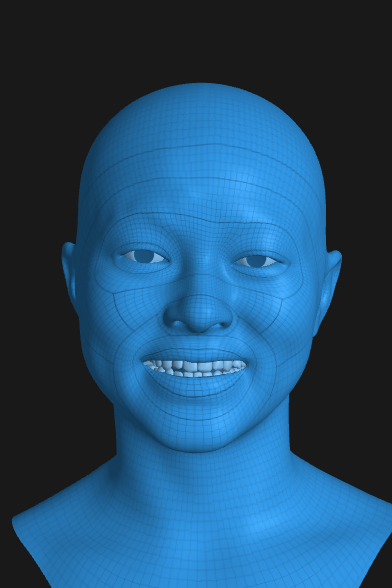}
    \\
  \includegraphics[height=\height]{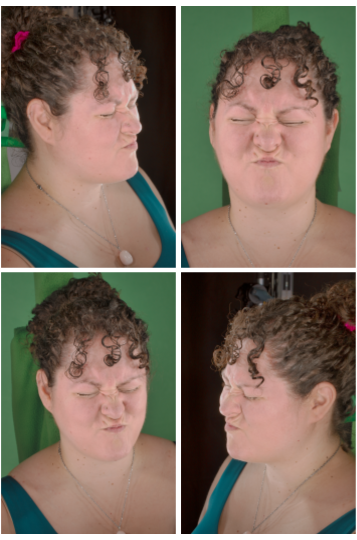}  
  & \includegraphics[height=\height]{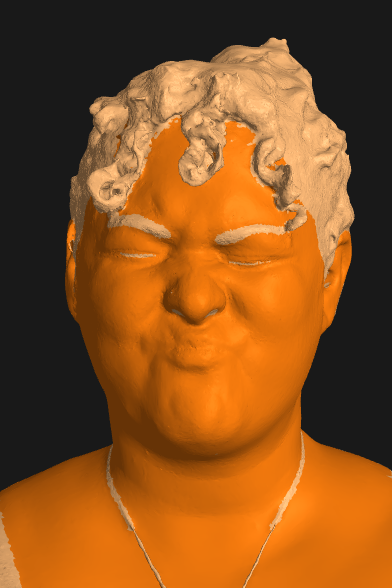}  
  & \includegraphics[height=\height]{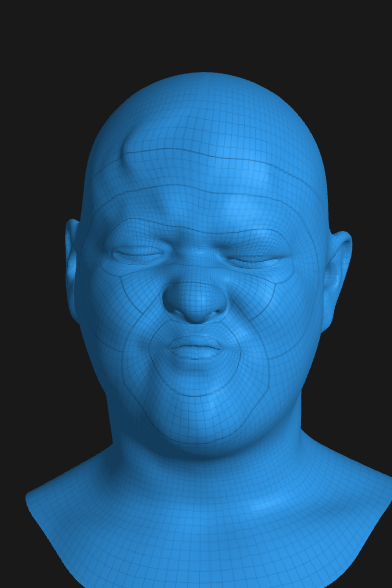} 
  & \includegraphics[height=\height]{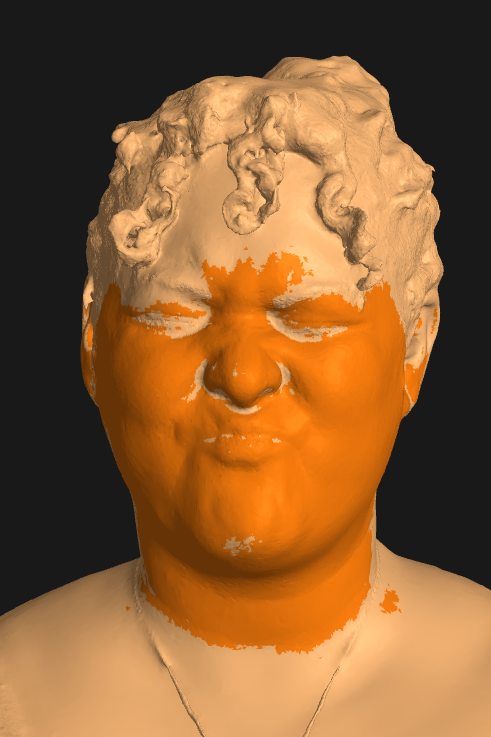} 
  & \includegraphics[height=\height]{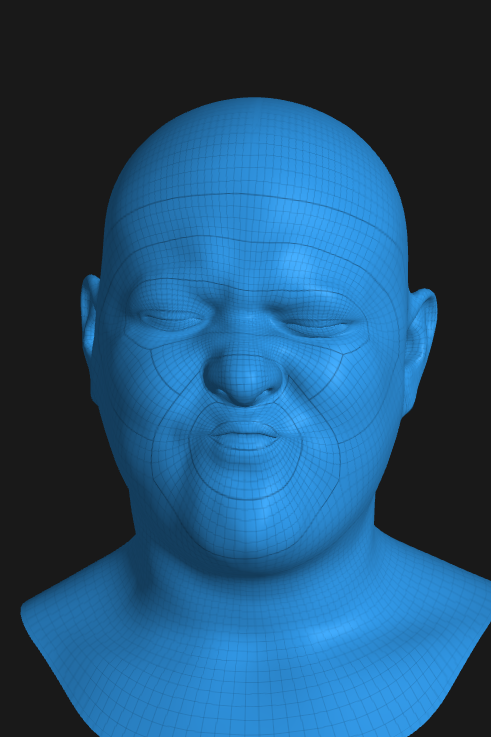} 
  & \includegraphics[height=\height]{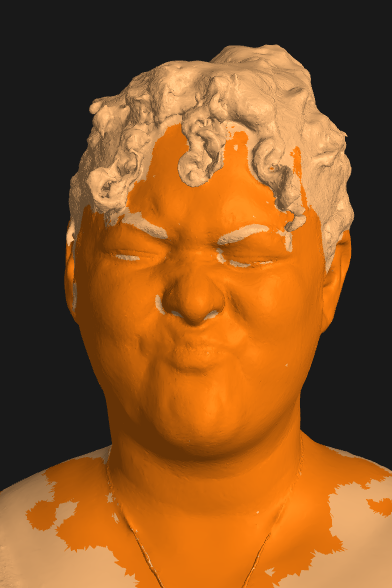}
 & \includegraphics[height=\height]{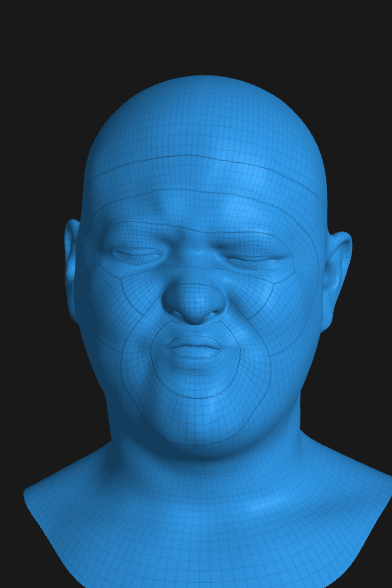}
    \\
  \includegraphics[height=\height]{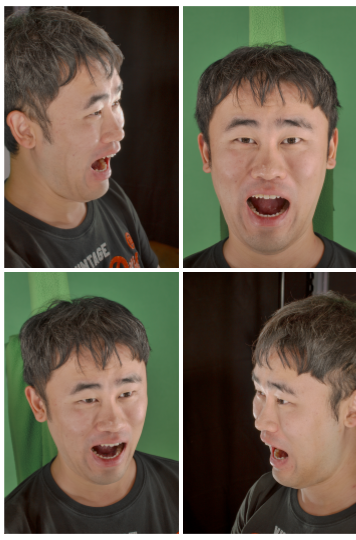}  
  & \includegraphics[height=\height]{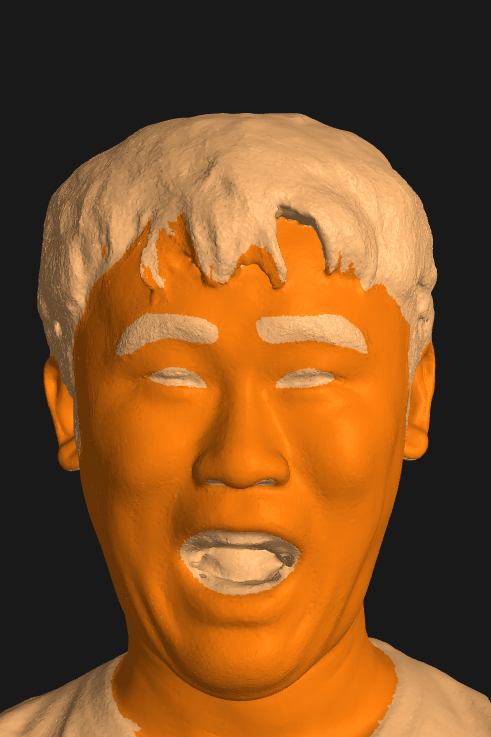}  
  & \includegraphics[height=\height]{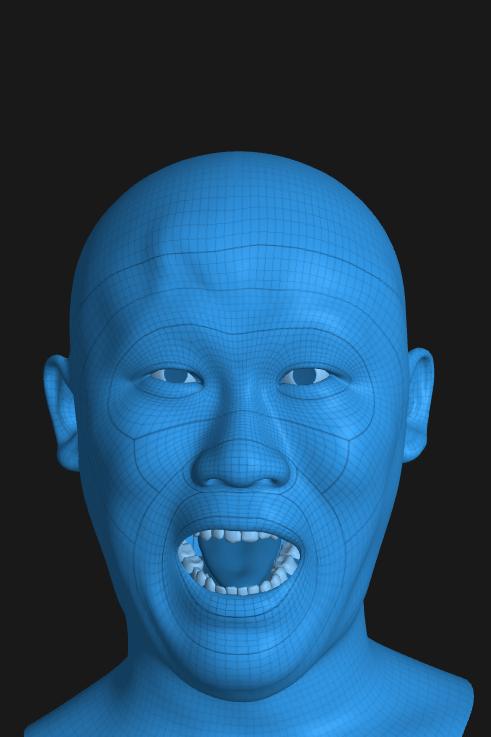} 
  & \includegraphics[height=\height]{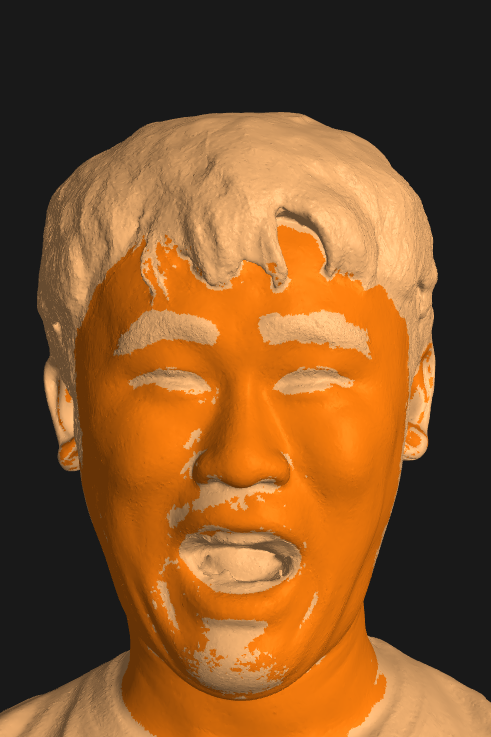} 
  & \includegraphics[height=\height]{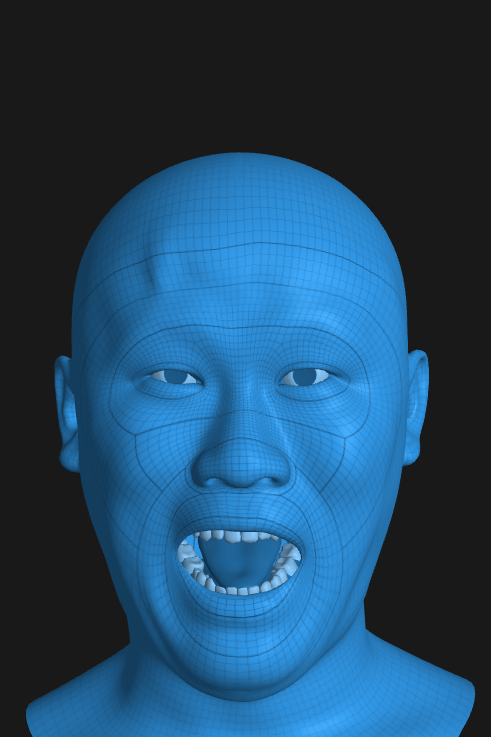} 
  & \includegraphics[height=\height]{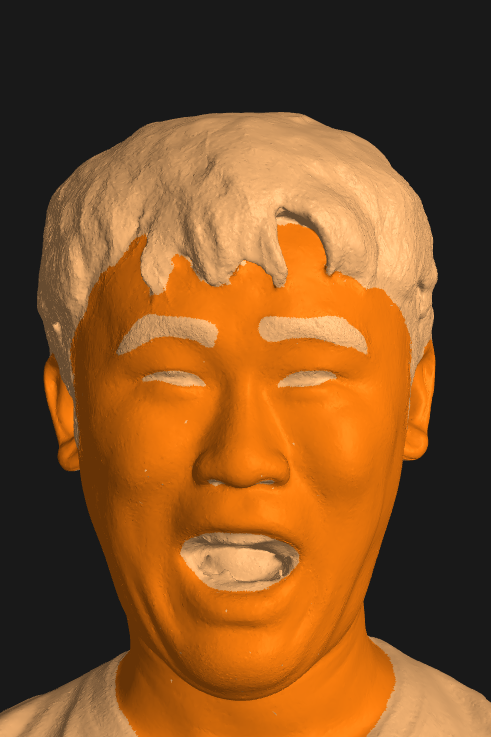}
 & \includegraphics[height=\height]{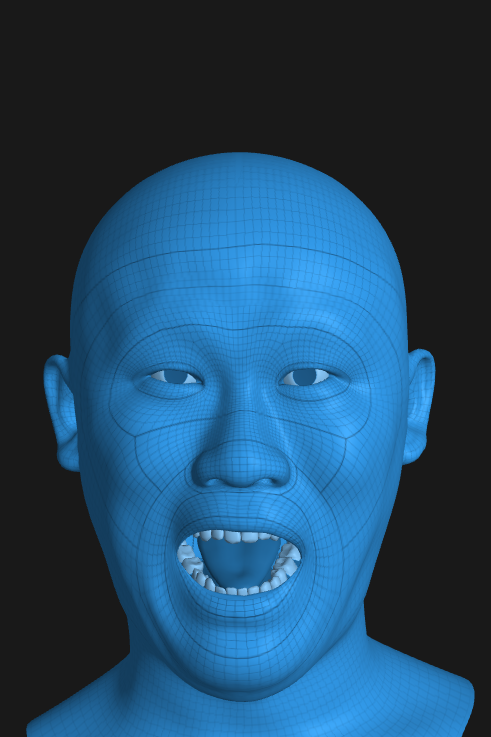}
    \\ 
RGB Images & Segmentation & Registration & Segmentation & Registration & Segmentation & Registration \\
& \multicolumn{2}{c}{2D~\cite{lugaresi2019mediapipe}} & \multicolumn{2}{c}{3D~\cite{sharp2022diffusionnet}} & \multicolumn{2}{c}{\textbf{Ours}}
\end{tabular}
\end{center}
\caption{We compare our method with the 2D baseline (MediaPipe~\cite{lugaresi2019mediapipe}) and the best performing 3D baseline (DiffusionNet~\cite{sharp2022diffusionnet}) on the real data. 
%
We evaluate both the segmentation and registration quality of 2D and 3D baselines against ours. Dark orange is the predicted skin. Our method outperforms pure 2D and 3D methods. More results on real data are in  Sup. Mat.}
\label{fig:qualitative-comparison}
\end{figure*}

We compare our method to a 2D segmentation baseline from MediaPipe~\cite{lugaresi2019mediapipe} and a 3D segmentation baseline using Diffusionnet~\cite{sharp2022diffusionnet}. 
For DiffusionNet, we experimented with a range of input features and only report the combinations which have made notable improvements compared to solely using 3D Euclidean coordinates. Point-to-surface distance $d_{\text{surface}}$ and mIoU reported in Tab.~\ref{tab:quantitative-comparison} demonstrate that our model outperforms both 2D and 3D baselines on real and synthetic data. 
In Fig.~\ref{fig:qualitative-comparison}, both the segmentation and registration results on the real data show that our model produces a segmentation mask that leads to a more accurate head geometry. On the contrary, the 2D segmenter leads to unwanted bumps on forehead due to artifacts on the skin mask, while the 3D segmenter filters out more skin regions than necessary and also fails to remove artifacts near the skin, as shown in the last row of Fig.~\ref{fig:qualitative-comparison}.

\vspace{-0.2cm}

\begin{table}[tb]
  \centering
  \caption{\textbf{Quantitative results.} For both segmentation (mIoU) and downstream registration (point-to-surface distance $d_{\text{surface}}$), our model outperforms the 2D and 3D baselines on synthetic and real data. Note that $d_{\text{surface}}$ is reported in millimeters (mm). For DiffusionNet (3D model), we used a combination of heat kernel signatures (HKS), surface variation ($\sigma_{30}$), and vertex color.
  }
  \vspace{-7pt}
   \resizebox{0.9\columnwidth}{!}{
        \begin{tabular}{lcccc}
        \toprule
        & \multicolumn{2}{c}{\textbf{Synthetics}} & \multicolumn{2}{c}{\textbf{Real}} \\
        & mIoU $\uparrow$ & $d_{\text{surface}}$ $\downarrow$  & mIoU $\uparrow$ & $d_{\text{surface}}$ $\downarrow$\\
        \midrule
        2D: Baseline                      & 0.786 &  0.427 $\pm$ 0.72 &  0.687  & 0.510 $\pm$ 0.50 \\
        \midrule
        3D: hks                           & 0.253 & 1.762 $\pm$ 5.03  &  0.089   &   2.05 $\pm$ 0.74  \\
        3D: hks, color                    & 0.670 & 0.430 $\pm$ 0.81  &  0.483   &  1.75 $\pm$ 0.77  \\
        3D: hks, $\sigma_{30}$            & 0.608 & 1.554 $\pm$ 3.61  &  0.499   &  1.31 $\pm$ 0.74  \\
        3D: hks, color, $\sigma_{30}$     & 0.749 & 0.454 $\pm$ 0.91  &  0.590   &  0.881 $\pm$ 0.48 \\
        \midrule
        Ours                 &  \textbf{0.923} & \textbf{0.328 $\pm$ 0.29}  &  \textbf{ 0.704}  &  \textbf{0.364 $\pm$ 0.32} \\
        \bottomrule 
        \end{tabular}%
   }
  \label{tab:quantitative-comparison}%

\vspace{-0.2cm}
\end{table}

\subsection{Ablation Study}

\paragraph*{Architecture.}
Tab.~\ref{tab:models-comparison} compares two feature extractors (DinoV2 and SAM), as well as two 3D network architectures (DiffusionNet and vanilla per-vertex MLP).
Using DinoV2 or SAM ViT backbones yields comparable segmentation results with negligible impact on downstream registration quality.
%
DiffusionNet improves upon vanilla MLP by approximately $0.9\%$ on mIoU and $d_{\text{surface}}$. 
%

\paragraph*{Projection.}
Tab.~\ref{tab:models-comparison} shows that considering vertex visibility further improves segmentation and thus registration on both synthetic and real scans.
%
Variance of image features across views helps the model to identify ambiguous regions on the scan mesh (see Sup. Mat.).

\paragraph*{Geometric Features.}
By including surface variation ($\sigma_{30}$) and using DiffusionNet, we achieve the best results on both real and synthetic scan meshes.
$\sigma_{30}$ supplement the image features with important information about the geometry such as the noise level of the surface, especially for those partially observed vertices where ViT features may be insufficient.
%
Meanwhile, HKS did not capture local geometric properties, which does not improve model performance.

\begin{table}[tb]
  \centering
  \caption{\textbf{Ablation study.} We compare different network architectures, DinoV2 and SAM, MLP and DiffusionNet~(Diff.). For projection methods, we show mean~$\mu$, variance~$\sigma^2$, and visibility-awareness~$\bar{\mu}, \bar{\sigma}^2$. 
  We compare Heat Kernel Signature~(HKS) and surface variation~$\sigma_{30}$ as augmenting geometric features.
}
  \vspace{-7pt}
   \resizebox{0.9\columnwidth}{!}{
    \begin{tabular}{lcccc}
    \toprule
    & \multicolumn{2}{c}{\textbf{Synthetics}} & \multicolumn{2}{c}{\textbf{Real}} \\
    & mIoU $\uparrow$ & $d_{\text{surface}}$ $\downarrow$  & mIoU $\uparrow$ & $d_{\text{surface}}$ $\downarrow$\\
    \midrule
    \multicolumn{1}{l}{\textbf{Architecture}} \\
    \midrule
    SAM + mlp                    & 0.899 & 0.395 $\pm$ 0.58  & 0.699  & 0.514 $\pm$ 0.42   \\
    SAM + Diff.                  & 0.899 & 0.395 $\pm$ 0.56  & 0.700  & 0.483 $\pm$ 0.58  \\
    DinoV2 + mlp                 & 0.906 & 0.392 $\pm$ 0.57  & 0.692  & 0.497 $\pm$ 0.59 \\
    DinoV2 + Diff.               & 0.911  & 0.389 $\pm$ 0.56 &  0.693  & 0.477 $\pm$ 0.61    \\ 
    \midrule
    \multicolumn{1}{l}{\textbf{Projection}} \\
    \midrule
    $\mu$                                              & 0.906   & 0.396 $\pm$ 0.57 &  0.699    &  0.533 $\pm$ 0.59   \\
    $\mu$, $\sigma^2$                                  & 0.911   & 0.389 $\pm$ 0.56 &  0.694    &  0.477 $\pm$ 0.61   \\
    (visibility-aware) $\bar{\mu}$                     & 0.912   & 0.391 $\pm$ 0.58 &  0.709    &  0.464 $\pm$ 0.63   \\
    (visibility-aware) $\bar{\mu}$, $\bar{\sigma}^2$   & 0.917   & 0.391 $\pm$ 0.56 & \textbf{0.712} &  0.479 $\pm$ 0.61   \\
    \midrule
    \multicolumn{1}{l}{\textbf{Geometric Features}} \\
    \midrule
    DinoV2 + HKS                       &0.918           & 0.396 $\pm$ 0.58  &   0.698  &  0.476 $\pm$ 0.60 \\
    DinoV2 + $\sigma_{30}$             & \textbf{0.923} & \textbf{0.328 $\pm$ 0.29}  &   0.704  &  0.364 $\pm$ 0.32 \\
    DinoV2 + $\sigma_{30}$ + HKS       &0.921           & 0.332 $\pm$ 0.30  &   0.701  &  \textbf{0.331} $\pm$ 0.33 \\
    \bottomrule 
    \end{tabular}%
  }
  \label{tab:models-comparison}%
\end{table}%

\vspace{-0.2cm}

\section{Conclusion}

In this paper, we introduce a novel 3D head scan segmentation algorithm that outputs high-quality vertex-level segmentation, which leads to a more robust registration. We propose to lift 2D image features into the 3D space and learn a DiffusionNet-based 3D segmentation model, which is further augmented with geometric features. Our method outperforms pure 2D and 3D methods on both segmentation and downstream registration. %
Regarding limitations, our method only outputs binary classification of skin and non-skin, which limits its capability in ambiguous regions like short stubble facial hair. Regressing the distance to the underlying surface can potentially further improve registration quality.
%




\bibliographystyle{eg-alpha-doi} 
\bibliography{egbibsample}       


\newpage


\end{document}